# Solar System Ice Giants: Exoplanets in our Backyard.

(Cover page)

Co-authors and endorsers:

**Abigail Rymer**[1] (JHUAPL, 11101 Johns Hopkins Road, Laurel 20723, USA, +1 443-778-2736, abigail.rymer@jhuapl.edu)

Kathleen Mandt[1], Dana Hurley[1], Carey Lisse[1], Noam Izenberg[1], H.Todd Smith[1], Joseph Westlake[1], Emma Bunce[2], Christopher Arridge[3], Adam Masters[4], Mark Hofstadter[5], Amy Simon[6], Pontus Brandt[1], George Clark[1], Ian Cohen[1], Robert Allen[1], Sarah Vine[1], Kenneth Hansen[7], George Hospodarsky[8], William Kurth[8], Paul Romani[6], Laurent Lamy[9], Philippe Zarka[9], Hao Cao[10], Carol Paty[11], Matthew Hedman[12], Elias Roussos[13], Dale Cruikshank[14], William Farrell[6], Paul Fieseler[5], Andrew Coates[15], Roger Yelle[16], Christopher Parkinson[7], Burkhard Militzer[17], Denis Grodent[18], Peter Kollmann[1], Ralph McNutt[1], Nicolas André[19], Nathan Strange[5], Jason Barnes[20], Luke Dones[21], Tilmann Denk[22], Julie Rathbun[23], Jonathan Lunine[12], Ravi Desai[4], Corey Cochrane[5], Kunio M. Sayanagi[24], Frank Postberg[25], Robert Ebert[21], Thomas Hill[26], Ingo Mueller-Wodarg[4], Leonardo Regoli[7], Duane Pontius[27], Sabine Stanley[1,49], Thomas Greathouse[21], Joachim Saur[28], Essam Marouf[29], Jan Bergman[30], Chuck Higgins[31], Robert Johnson[32], Michelle Thomsen[23], Krista Soderlund[33], Xianzhe Jia[7], Robert Wilson[34], Jacob Englander[6], Jim Burch[21], Tom Nordheim[5], Cesare Grava[21], Kevin Baines[5], Lynnae Quick[35], Christopher Russell[36], Thomas Cravens[37], Baptiste Cecconi[9], Shahid Aslam[6], Veronica Bray[16], Katherine Garcia-Sage[6,38], John Richardson[39], John Clark[40], Sean Hsu[34], Richard Achterberg[6,41], Nick Sergis[42], Flora Paganelli[43], Sasha Kempf[34], Glenn Orton[5], Ganna Portyankina[34], Geraint Jones[15], Thanasis Economou[44], Timothy Livengood[6], Stamatios Krimigis[1,42], James Szalay[45], Catriona Jackman[46], Phillip Valek[21], Alain Lecacheux[9], Joshua Colwell[47], Jamie Jasinski[5], Federico Tosi[48], Ali Sulaiman[8], Marina Galand[4], Anna Kotova[13], Krishan Khurana[36], Margaret Kivelson[36], Darrell Strobel[49], Aikaterina Radiota[18], Paul Estrada[43], Stefano Livi[21], Abigail Azari[7], Japheth Yates[50], Frederic Allegrini[21], Marissa Vogt[40], Marianna Felici[40], Janet Luhmann[51], Gianrico Filacchione[48], Luke Moore[48].

[1]Johns Hopkins University Applied Physics Laboratory, [2]University of Leicester, UK [3]Lancaster University, UK, [4]Imperial College, UK, [5]Jet Propulsion Laboratory, California Institute of Technology, [6]NASA Goddard Space Flight Center, [7]University of Michigan, [8]University of Iowa, [9]LESIA, Observatoire de Paris, France, [10]Harvard University, [11]Georgia Institute of Technology, [12]Cornell, [13]Max Planck Institute, Germany, [14]NASA Ames, [15]MSSL, University College London, UK, [16]University of Arizona, [17]University of California, Berkeley, [18]Université de Liège, Belgium, [19]Research Institute in Astrophysics and Planetology, France, [20]University of Idaho, [21]Southwest Research Institute, [22]Freie Universität Berlin, Germany, [23]PSI, [24]Hampton University, [25]University of Heidelberg, Germany,[26]Rice University, [27]Birmingham-Southern College, [28]University of Cologne, Germany, [29]San Jose State University, [30]Swedish Institute of Space Physics, [31]Middle Tennessee State University, [32]University of Virginia, [33]University of Texas at Austin, [34]University of Colorado, [35]Smithsonian Institute, CEPS, [36]University of California, Los Angeles, [37]Kansas University, [38]Catholic University of America, [39]MIT, [40]Boston University, [41]University of Maryland, [42]Academy of Athens, Greece, [43]SETI Institute, [44]University of Chicago, [45]Princeton, [46]University of Southampton, UK, [47]University of Central Florida, [48]INAF – IAPS, Italy, [49]Johns Hopkins University, [50]ESA, [51]University of California, Berkeley.

This White Paper is endorsed by >100 scientists and engineers from the USA and Europe, many of whom are early career scientists representing the driving force of the heliophysics community in the decades to come.

**Motivation.**
Future remote sensing of exoplanets will be enhanced by a thorough investigation of our solar system Ice Giants (Neptune-size planets). What can the configuration of the magnetic field tell us (remotely) about the interior, and what implications does that field have for the structure of the magnetosphere; energy input into the atmosphere, and surface geophysics (for example surface weathering of satellites that might harbour sub-surface oceans). How can monitoring of auroral emission help inform future remote observations of emission from exoplanets? **Our Solar System provides the only laboratory in which we can perform in-situ experiments to understand exoplanet formation, dynamos, systems and magnetospheres.**

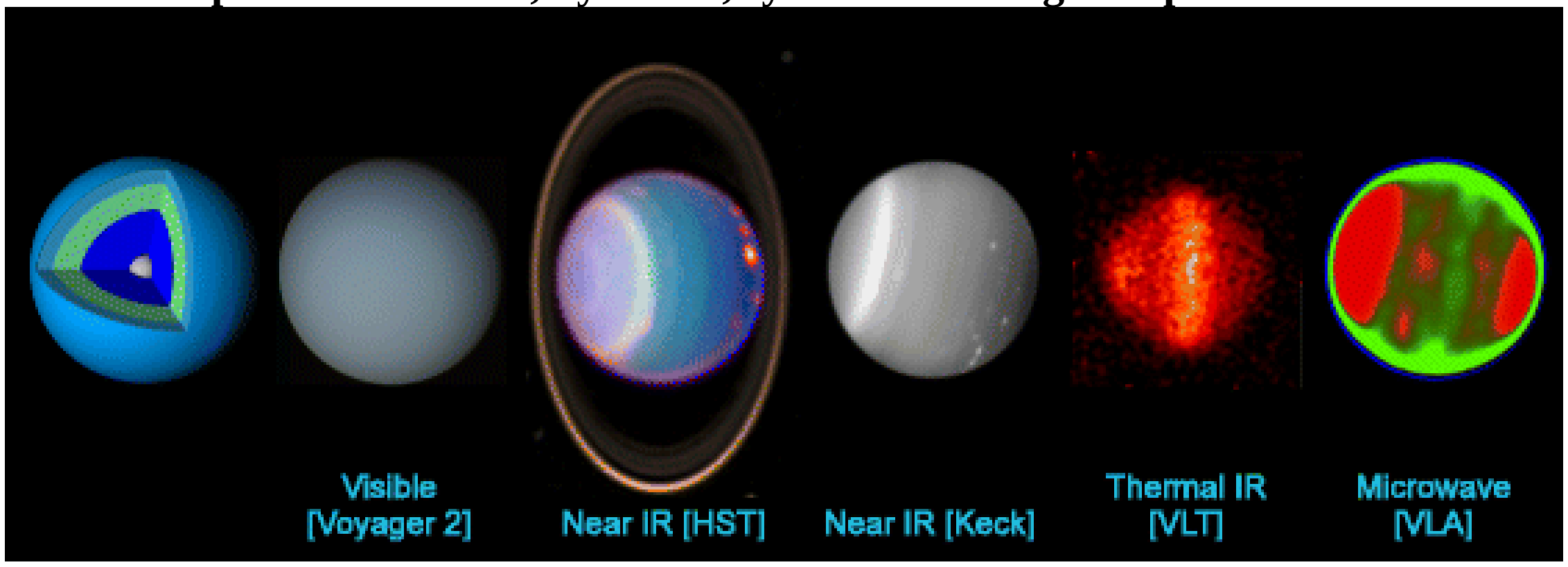

**Figure 1.** A notional model of Uranus' interior (far left, with a rock core surrounded by ionic and normal water oceans [blue and green, respectively] and a gaseous outer later primarily composed of hydrogen and helium). The 5 images on the right show Uranus' appearance at multiple wavelengths, with wavelength and instrument indicated in the figure. Although Uranus appeared relatively tranquil in images obtained by Voyager 2 due to obscuring tropospheric hazes prevalent over the summer pole, imaging at longer wavelengths and other seasons demonstrate the wide range of discrete cloud activity and the distributions of gaseous opacity sources on the Ice Giant. Credits: (a) NASA/JPL; (b) E. Karkoschka (University of Arizona, USA), Hubble Space Telescope and NASA; (c) H. Hammel (Space Science Institute, Boulder, USA), I. de Pater (University of California Berkeley, USA), W.M. Keck Observatory; (d) G. Orton (NASA JPL); (e) M. Hofstadter (NASA JPL). From Arridge et al., [2011].

'Ice giants' are the only major category of Solar System object never to have had a dedicated mission and represent one of the largest groups of detected exoplanets [Fulton et al., 2017]. We know very little about our own ice giants, and the potential science return from a Galileo- or Cassini-class mission to Uranus and/or Neptune is immense. Here we outline the science case for exploration of an Ice Giant in furthering our understanding of this important class of exoplanets. White Papers submitted to the Planetary Science Decadal Survey 2013-2023 [Hofstadter et al., 2010] and the Heliophysics Science Decadal Survey 2013-2023 [Rymer et al., 2010; Hess et al., 2010] provide a persuasive case for an Ice Giant (Uranus or Neptune) orbiter to investigate the composition, structure, atmosphere and internal dynamo of ice giants and the nature and stability of their moon and ring systems. Both resultant decadal surveys advocate a future mission to these solar system targets.

Exploration of at least one ice giant system is critical to advance our understanding of the Solar System, exoplanetary systems, and to advance our understanding of planetary formation and evolution. Three key points highlight the importance of sending a mission to our ice giants, Uranus and Neptune. First, they represent a class of planet that is not well understood, and which is fundamentally different from the gas giants (Jupiter and Saturn) and the terrestrial planets. Ice giants are, by mass, about 65% water and other so-called "ices," such as methane and ammonia. In spite of the "ice" name, these species are thought to exist primarily in a massive, super-critical liquid water ocean. No current model for their interior structure is consistent with all observations. A second key factor in their importance is that ice giants are extremely common in our galaxy. They are much more abundant than gas giants such as Jupiter. Exploration of our local ice giants would allow us to better characterize exoplanets. The final point to emphasize about ice giants is that they challenge our understanding of planetary formation, evolution, physics and chemistry.

**Progress Since the New Worlds New Horizons Decadal Survey.**
Within the past decade it has been realized that Neptune-size planets are among the most common class of exoplanet in our galaxy, [Fulton et al., 2017].

**Areas Where Significant Progress will Likely be made with Current and Upcoming Ground- and Space-Based Facilities.**
Radio-emissions from exoplanets might be detectable. The relatively high contrast between planetary and solar low-frequency radio emissions suggests that the low-frequency radio range may be well adapted to the direct detection of exoplanets. Zarka et al., [2007] review the most significant properties of planetary radio emissions (auroral as well as satellite induced) and show that their primary engine is the interaction of a plasma flow with an obstacle in the presence of a strong magnetic field (of the flow or of the obstacle). Extrapolating this scaling law to the case of exoplanets, they find that hot Jupiters may produce very intense radio emissions due to either magnetospheric interaction with a strong stellar wind or to unipolar interaction between the planet and a magnetic star (or strongly magnetized regions of the stellar surface). In the former case, similar to the magnetosphere–solar wind interactions in our solar system or to the Ganymede–Jupiter interaction, a hecto-decameter emission is expected in the vicinity of the planet with an intensity possibly $10^3$–$10^5$ times that of Jupiter's low frequency radio emissions. In the latter case, which is a giant analogy of the Io–Jupiter system, emission in the decameter-to-meter wavelength range near the footprints of the star's magnetic field lines interacting with the planet may reach 106 times that of Jupiter (unless some "saturation" mechanism occurs). A hot spot was already tentatively detected in visible light near the sub-planetary point in the system of HD179949. These emissions vary with host star activity – a very exciting element for future study.

**Exoplanet Investigations Enabled by Planetary Missions.**
In situ study of an Ice Giant will enable numerous investigations that, despite several decades of study, are still not fully understood. These include the following top-level questions that have direct relevance to exoplanets:

**1. Auroral configuration and emission.**

Auroral emissions are generated above the ionosphere at kilometric (radio) wavelengths (1–1,000 kHz) (known as Uranus Kilometric Radiation—UKR). As at other planets (e.g. Jupiter, Figure 2) UKR is thought to be generated by the Cyclotron Maser Instability (CMI) around the magnetic poles and therefore is a remote marker of planetary rotation. A dayside spot-like transient near-equinoctial was recently observed at Uranus a different type of emission than at solstice and in turn witness a different solar wind/magnetosphere interaction which dramatically evolves along the planetary revolution [Lamy et al., 2017]. UKR displays a rich variety of components characteristic of Ice Giants, including unique features such as time-stationary radio sources [Zarka et al., 1987; Desch et al., 1991].

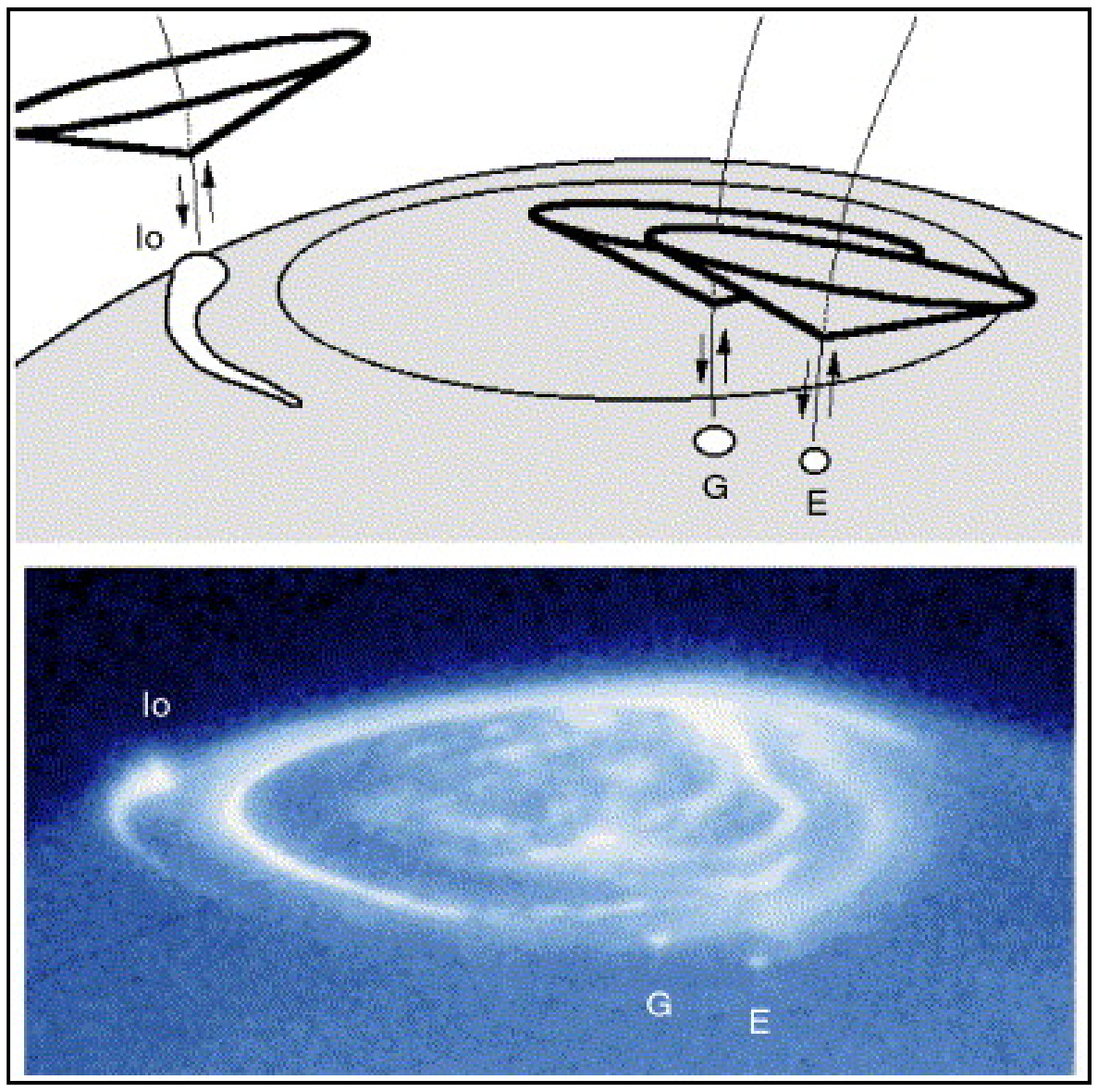

Fig 2. (Bottom) HST UV image of northern Jovian auroral regions, showing clearly the bright main auroral oval and the footprints of Io (plus a tail-like structure), Ganymede and Europa flux tubes. Courtesy: R. Prangé , L. Pallier, and J.T. Clarke. (Top) Sketch of radio emission hollow conical beams produced above the UV hot spots by the energetic electrons precipitated along the satellite flux tubes or reflected upwards by magnetic mirroring. [Zarka et al., 2007]

Understanding the circumstances under which planetary radio emissions are generated is of prime importance for using them to detect exoplanetary magnetic fields [Farrell et al., 1999; Zarka et al., 2007] (important for the development and protection of atmospheres and life). **Unlike our Solar System, eccentric and complex orbital characteristics appear to be common in other planetary systems, so that the understanding of radio emission produced by Uranus could have profound importance for interpreting future radio detections of exoplanets.**

**2. Magnetospheric Transport/Atmospheric Energy Deposition.**

The peculiar combination of magnetic and spin axes at both Uranus and Neptune means that the plasma sheet is twisted as the planet rotates causing magnetic field lines in the roughly cylindrical magnetotail to be wound into a helical (corkscrew) shape [*Hill et al.*, 1983]. **Mechanisms for plasma transport and diffusion, that are well understood at other planets, have never been studied in this type of geometry.** What is the influence of atmospheric composition and temperature on magnetosphere-ionosphere coupling processes that govern convection and auroral processes?

The unique feature of the Voyager 2 encounter was the fact that the spin axis of Uranus was aligned nearly along the planet-sun line. This led to the condition that a solar-wind-driven magnetospheric convection system was effectively decoupled from corotation, as noted above. Stated another way, the flow system rotational electric field, which ordinarily would have "shielded" the middle magnetosphere from the solar wind, was oriented in such a way that solar wind effects could penetrate deeply into the magnetosphere. The consequences included:

a) Convection patterns similar to Earth's with a well-defined plasma pause.

b) Strong dynamics including Earthlike injection phenomena;
c) An electron radiation belt that was as intense as the most intense seen at Earth; and
d) The strongest whistler-mode emissions seen at any of the outer planets.

**3. Radiation Belts (Energetic Particle Trapping).**
One might expect that the configuration at Uranus would lead to less efficient particle trapping and heating required to form radiation belts. In fact, Voyager 2 found electron radiation belts at Uranus of intensity similar to those at Earth and much more intense than those at Saturn [Krimigis et al., 1986]. The ion radiation belts are similar between Uranus and Saturn, although they differ in composition. **How stable are the Uranian radiation belts? Are they always present? Can we guide the search for exoplanets with magnetic fields by identifying which of them have radiation belts?**

**4. Bulk Composition and Internal Structure.**
Composition and structure are the properties that define ice giants as distinct from terrestrial and gas giant planets. Knowledge of the ice-to-rock and ice-to-gas ratios as well as the absolute abundance of certain key species, such as noble gases and water, tells us about conditions in the planetary nebula and the planet formation process [*Hersant et al.*, 2004]. Whether the gas and heavier components are segregated or well mixed today offers additional clues as to how and when each component was incorporated into the planet, and how much mixing occurred. That mixing strongly influences the chemical and thermal evolution of the planet. Knowledge of the bulk composition and interior structure also allows us to relate current observed properties of the atmosphere (abundances of trace or disequilibrium species such as $NH_3$ or CO, and the temperature profile) to details of the heat flow, convection, chemistry and dynamo action occurring today at depth. **Understanding the composition and structure of our Solar System's ice giants is a necessary prerequisite to identifying them around other stars from the minimal information available (such as mass and radius), and recognizing if those exoplanetary systems contain a type of planet not seen in our Solar System.**

**5. Intrinsic magnetic field.**
The ice giants' multipolar, non-axisymmetric magnetic fields were a surprise upon their discovery, and it is still not understood why these bodies generate remarkably different fields compared to all other planets in our solar system, whose intrinsic fields are dipole-dominated and nearly aligned with their rotation axes [e.g., Schubert and Soderlund, 2011]. **By understanding the dynamos of our solar system, we would be able to predict the magnetic field strengths and morphologies of exoplanetary dynamos with more confidence as well [e.g., Tian and Stanley 2013].** The mission could answer key questions that characterize intrinsic magnetic fields and constrain the dynamo processes responsible for their generation: What is the configuration of Uranus' intrinsic magnetic fields? Has secular variation occurred since the Voyager 2 observations? What is the rotation rate of the bulk interior and how does it compare to the radio rotation rate?

**Measurement Requirements.**
The 2013-2023 NASA Planetary Decadal survey and a recent NASA mission study (Hofstadter et al. 2017) describe the science drivers for several measurements. We advocate in particular for high resolution magnetometry (like the Cassini MAG), microwave sounding, multi-wavelength imaging spectroscopy, ion plasma composition and full electron pitch angle distributions across the widest possible dynamic range (a combination sensor akin to the JEDI-JADE plasma suite on the Juno spacecraft might be most appropriate), radio and plasma wave package with similar capabilities to those onboard the Cassini (RPWS) and Juno (WAVES) spacecraft, neutral particles and dust detectors (something like the Cassini INMS and CDA instruments), an atmospheric entry probe to measure noble gases and isotopic ratios, radio science Ka-band transponder along with laboratory and ground based support measurements. An ENA camera like Cassini-INCA would make important measurements and also provide opportunities for heliophysics observations with cross-discipline relevance. Significant science payloads could be inserted into orbit around Uranus using chemical propulsion alone using relatively modest launch vehicles. Thus cost is the single biggest factor limiting instrument payload size making cost sharing across disciplines and internationally a very attractive option in producing a feasible cost-effective mission. More broadly we advocate that this mission be considered in concert or even coupled with an Interstellar Probe mission to explore the distant solar system and beyond, as well as the potential to enhance science return with the use of small satellites and cube-satellites.

We further endorse that the community endeavour to perform such a mission in collaboration with the Heliophysics and Planetary Science NASA divisions as well as the international community.

A final note, while the science outlined here is motivated by our desire to make un-paralleled magnetospheric measurements at Uranus, the tour phase out to 20 AU will afford a rare opportunity to make solar wind observations in the outer heliosphere and as such appeal to an even broader section of the heliophysics community.